\documentstyle[prb,aps]{revtex}
\begin{document}
\widetext
\draft
\title
{Ionization degree of the electron-hole plasma in semiconductor quantum wells}
\author{M. E. Portnoi\cite{byline} and I. Galbraith} 
\address{
Physics Department, Heriot-Watt University, Edinburgh EH14 4AS, 
United Kingdom} 


\maketitle

\begin{abstract}

The degree of ionization of a nondegenerate two-dimensional 
electron-hole plasma is calculated using the modified law of 
mass action, which takes into account {\it all} bound and 
unbound states in a screened Coulomb potential.   
Application of the variable phase method to this potential allows 
us to treat scattering and bound states on the same footing.
Inclusion of the scattering states leads to a strong deviation from 
the standard law of mass action. 
A qualitative difference between mid- and wide-gap semiconductors 
is demonstrated. 
For wide-gap semiconductors at room temperature, 
when the bare exciton binding energy is of the order of $~k_BT$, 
the equilibrium consists of an almost equal mixture of correlated 
electron-hole pairs and uncorrelated free carriers.

\end{abstract}


\section{INTRODUCTION}
\label{intro}

The drive for ever higher storage capacity has led to the development
of semiconductor lasers operating in the blue spectral region, 
based on ZnSe\cite{Sony} and GaN.\cite{Nichia}
Along with the large energy gap of these materials comes a large exciton 
binding energy, of the same order as $~k_BT~$ at room temperature.
As has been well known since the seventies,\cite{Excigain} excitonic 
gain processes are important in wide-gap semiconductors, and 
their importance is further enhanced in quantum well 
structures where the binding energy may be considerably larger than 
$~k_B T$ (e.g., $\sim$35 meV in ZnCdSe/ZnSe quantum wells).

Theoretical treatments of GaAs and InP based lasers are well established 
using  a microscopic many-body approach based on linear response 
theory.\cite{LaserBook} 
Screening and band-gap renormalization effects are included assuming the 
injected carriers form a completely ionized electron-hole plasma. 
Such treatments have been successful in explaining many of the observed 
features of mid-infrared laser diodes. 
Complex valence band effects and strain effects as well as carrier 
thermalization effects have all been included at various levels of 
complexity. 
In this way, a relatively complete understanding exists for the basic
operation of these lasers.  

In wide-gap semiconductors however, the strong Coulomb interaction leads 
to the existence of bound exciton states, which persist even at 
elevated densities and temperatures. 
As such, the conventional assumption that the inversion is in the form of 
an electron-hole plasma with no excitons present deserves closer examination. 
A self-consistent description where both bound and unbound states 
are treated on an equal footing is required.
Unfortunately, as far as we are aware no comprehensive theoretical 
treatment of this problem exists. 
Treatments based on bosonic exciton operators have been proposed\cite{HH77} 
but this approach breaks down at high injection when the screening of the 
Coulomb potential weakens the binding and produces a population of unbound 
scattering states, which  clearly  do not  exhibit bosonic character. 
On the other hand, a treatment based around fermionic electron and hole 
operators is complex when higher order excitonic correlations are important.
\cite{Lindberg,Pereira}
A natural concept in considering this issue is the degree of ionization 
in the interacting electron-hole plasma, and in this paper we calculate 
this for a two-dimensional (2D) plasma.
We focus on 2D for two reasons.
Firstly, most modern semiconductor lasers are fabricated in quantum well
heterostructures.
Secondly, the presence of at least one bound state in the attractive 2D
potential requires a nonperturbative treatment of the screened Coulomb
interaction.

We will be mostly interested in the plasma properties induced by the pair 
Coulomb interaction between charged particles, 
neglecting band gap renormalization and phase-space filling effects, 
which have been extensively studied in both three-dimensional (3D) and 
2D cases.\cite{3Dplasma,2Dplasma}
These effects can be neglected only in the low-density (nondegenerate) 
limit, which is defined in 2D by the inequality
\begin{equation}
n\lambda_M^2/g~<~1~,
\label{ineq1}
\end{equation}
where $~n~$ is the 2D carrier density, $~g~$ is the spin degeneracy 
factor of 2D particles, 
and $~\lambda_M=(2\pi\hbar^2/M k_B T)^{1/2}~$ is the thermal wavelength.
For the two-component plasma the lighter carrier (usually electron) 
effective mass must be used to evaluate the thermal wavelength to ensure 
that condition (\ref{ineq1}) is valid for both types of carriers. 
Inequality (\ref{ineq1}) provides that the motion of excitons can also 
be considered as classical.
For GaAs at room temperature  
$~\lambda_{M_e}\,\approx\,1.66\times10^{-6}~{\rm cm}$, electron spin 
degeneracy $~g_e\,=\,2~$ and condition (\ref{ineq1}) is satisfied for 
$~n\,\alt\,7.2\times10^{11}~{\rm cm}^{-2}$.
The electron effective mass in wide-gap semiconductors is usually at
least two times larger than in mid-gap semiconductors, 
therefore condition (\ref{ineq1}) is valid over a wider range of 
carrier densities (e.g. for ZnSe at room temperature inequality 
(\ref{ineq1}) is satisfied for $~n\,\alt\,1.7\times10^{12}~{\rm cm}^{-2}$).
Thus, the nondegenerate (Boltzmann) limit is not only a convenient 
approximation, in which the Coulomb interaction is not hidden by the 
band-filling effects, but it also gives a realistic picture of 
the electron-hole plasma in wide-gap semiconductors at room temperature 
and moderate carrier densities. 
Lasing at anomalously low densities (below the Mott density) has been
reported in ZnCdSe/ZnSe quantum wells.\cite{Ding94}

Following an approach applied in 3D to nuclear matter,\cite{Ropke} 
an ionic plasma,\cite{Plasma86} and the electron-hole system in excited 
semiconductors\cite{Zimm} we divide the total electron (hole) density 
between two terms:  
\begin{equation}
n_a~=~n_a^0~+~n_a^{corr}~.
\label{eq01}
\end{equation}
The first term $~n_a^0~$ is the density of uncorrelated quasiparticles 
with renormalized energies.
This term is that part of the total density which is independent 
of the inter-particle interaction (see Appendix~\ref{BUformula}).  
All correlation effects both in the bound and continuum states are 
incorporated into the second term $~n_a^{corr}$, which is called the 
correlated density. 
The lower index in Eq.~(\ref{eq01}) is a species index, 
$~a=e~$ for electrons and $~a=h~$ for holes. 
It is also useful to introduce the degree of ionization of the 
electron-hole plasma
\begin{equation}
\label{alpha0}
\alpha~=~{n_e^0\over n_e}~=~{n_e^0\over n_e^0+n_e^{corr}},
\end{equation}
which characterizes the deviation of the thermodynamic properties of 
the electron-hole plasma from those of the ideal gas ($\alpha=1$). 
The knowledge of the degree of ionization is essential in determining 
the dominant lasing mechanism. 
When $~\alpha~$ is close to unity the main lasing mechanism is stimulated 
emission from the free-carrier plasma, for lower values of $~\alpha~$ 
several excitonic gain processes have to be considered.\cite{Ian96}

In the nondegenerate limit there is no need to go beyond two-particle 
correlations. 
This allows us to separate clearly the role of the inter-particle Coulomb 
interaction from the phase-space filling effects. 
In this limit,  the correlated and uncorrelated  densities are related by      
\begin{equation}  
n_a^{corr}~=~\sum_b n_a^0 n_b^0 
~{2\pi \beta \hbar^2 \over \mu_{ab}}~Z_{ab}~,
\label{eq02}
\end{equation}
where $~\beta=1/(k_BT)$, $~\mu_{ab}=M_a M_b/(M_a+M_b)~$ is the reduced 
effective mass, and $Z_{ab}$ is the two-body interaction part of the 
partition function. 
This relationship is derived in Appendix~\ref{BUformula}.
Note that due to charge-neutrality the total electron-hole density 
$~n_e=n_h=n~$ is independent of species, whereas $~n_e^0 \neq n_h^0~$ and 
$~n_e^{corr} \neq n_h^{corr}~$ if electrons and holes have different 
effective masses.
 
The electron-hole part of the partition function which exhibits bound 
states (excitons) is given  by 
\begin{eqnarray}
Z_{eh}&~=~&\sum_{m,\nu}\exp(-\beta E_{m,\nu}) 
\nonumber\\
&~+~&{1\over\pi}\,\int_0^\infty \left( \sum_{m=-\infty}^{\infty} 
{d\delta_m(k) \over dk} \right)\, 
\exp\left(-\beta{\hbar^2 k^2 \over 2\mu_{eh}}\right)\,dk~,
\label{eq03} 
\end {eqnarray}  
where $~m\hbar~$ is the projection of the angular 
momentum onto the axis normal to the plane of 2D motion 
($m=0, \pm 1, \pm 2, ~\ldots$), 
$~\hbar^2 k^2/(2\mu_{eh})~$ is the energy of the relative motion 
of the unbound (scattered) electron and hole, 
$\hbar k~$ is the magnitude  of the relative motion momentum, 
$~\delta_{m}(k)~$ are the 2D scattering phase shifts,\cite{SH67,LandauQM} 
$~E_{m,\nu}~$ are the bound-state energies 
(index $~\nu~$ enumerates bound states with given $~m$),
and the double sum in the first term ranges only over bound states. 
Equation (\ref{eq03}) is the 2D analogue of the Beth-Uhlenbeck 
formula\cite{BU37} and it can be derived in the same fashion as in the 
3D case, as shown in Appendix~\ref{BUformula}.
  
The scattering (integral) term in the right-hand side of Eq.~(\ref{eq03}) 
gives the contribution to $~Z_{eh}~$ of the continuum part of the energy 
spectrum. 
This term is usually neglected in calculations of the ionization degree 
of the electron-hole plasma.\cite{Excigain,Ian96,Cing96,BMBJAM98} 
In what follows we will show that at high enough temperature the scattering 
term is comparable to the bound-state sum and indeed this term has to be 
taken into account to ensure continuity of the partition function whenever 
bound states disappear with increasing screening.\cite{PGprb1}
 
The electron-electron and hole-hole parts of the partition function 
$~Z_{ee}~$ and $~Z_{hh}~$ contain the scattering term only. 
To calculate $~Z_{aa}~$ one must take into account the Pauli exclusion 
principle for identical particles, which modifies the sum over $~m$.
The electron-electron (hole-hole) part of the partition function is 
given by (see Appendix~\ref{BUformula})
\begin{equation}
Z_{aa}~=~{1 \over 2\pi} \sum_{m=-\infty}^{\infty}
\left \{2\,-\,(-1)^m \right\}
\int_0^\infty {d\delta_m(k) \over dk}
\exp\left(-{\beta}{\hbar^2 k^2 \over M_a}\right)\,dk~.
\label{eq04}
\end{equation}
Here we assume that both electron and hole states in quantum wells are 
two-fold degenerate. The only difference between $~Z_{hh}~$ 
and $~Z_{ee}~$ arises from the  difference between electron and hole  
effective masses. 
  
Equations (\ref{eq01})-(\ref{eq04}) provide a connection between 
the total electron-hole density $~n~$ and uncorrelated quasiparticle 
densities $~n_e^0~$ and $~n_h^0$. 
The  quasiparticle densities in turn govern 
the screening\cite{Zimm} and therefore the strength of interaction 
between particles, which defines uniquely the set of binding energies 
and scattering phase shifts which enter Eqs.~(\ref{eq03}) and 
(\ref{eq04}) for the two-body partition functions. 
These partition functions in turn define the ratio between 
$~n_a^0~$ and $~n_a^{corr}~$ via Eq.~(\ref{eq02}).
Thus, to find the degree of ionization of the electron-hole plasma 
one must solve the system of Eqs.~(\ref{eq01})-(\ref{eq04}) 
self-consistently, together with a reasonable model of the screened
interaction.

In the next section we discuss the statically screened Coulomb 
potential which we use to model the interaction between particles in 
an exciton/electron-hole plasma, and present results from the 
application of the variable phase method\cite{Calogero} to scattering 
and bound states in this potential. 
In Section~\ref{Results} we present and discuss the results of 
calculations of partition functions and the degree of ionization 
of the electron-hole plasma.
In Appendix~\ref{VPM} we derive the basic equations of the variable 
phase method, which is used for calculation of scattering phase shifts 
and binding energies.
The 2D analogue of the Beth-Uhlenbeck formula and the modified law 
of mass action are derived in Appendix~\ref{BUformula}.

\section{STATICALLY SCREENED COULOMB POTENTIAL}
\label{Coulomb}

There is an extensive literature dealing with different aspects of the 
screened Coulomb interaction in 2D systems.\cite{2Dscr}
In this paper we model this interaction by the well-known Thomas-Fermi
expression for a statically screened Coulomb potential \cite{SH67}
\begin{eqnarray}
V_s(\rho)~&=&~\mp {e^2 \over \epsilon} \int_{0}^{\infty}
{q J_0(q\rho) \over q+q_s}dq
\nonumber\\
&=&~\mp {e^2 \over \epsilon} 
\left \{{1 \over \rho}-{\pi \over 2} q_s [{\bf H}_0(q_s \rho)  
- N_0(q_s \rho)] \right\}~, 
\label{TF2D}
\end {eqnarray} 
where $~q_s~$ is the 2D screening wavenumber (which depends on 
temperature and carrier density), $~\epsilon~$ is the static 
dielectric constant of the semiconductor,
$~J_0(x)$, $~N_0(x)$, and $~{\bf H}_0(x)~$ are the Bessel, Neumann, 
and Struve functions respectively. 
The upper sign in Eq.~(\ref{TF2D}) is for electron-hole attraction,
the lower sign is for electron-electron or hole-hole repulsion.

Being the long-wavelength, static limit of the random phase approximation
for a purely 2D case,\cite{Stern67,HKbook} Eq.~(\ref{TF2D}) is the 
simplest model for the screened Coulomb potential in 2D.   
Nevertheless, this expression reflects the fact that the statically 
screened potential in 2D decreases at large distances slower 
than in the 3D case (as a power law rather than exponentially). 
Despite numerous realistic corrections,\cite{2Dscr,WBF89,Henriques} 
Eq.~(\ref{TF2D}) remains the most widely used approximation for the 
2D screening, especially for the screened exciton problem.
\cite{BMBJAM98,SpLM85,LSpM85,ES89,WKB97}
Optically active ($m=0$) bound states in the attractive, statically 
screened Coulomb potential [upper sign in Eq.~(\ref{TF2D})] have been  
studied using  a variational method,\cite{SpLM85,LSpM85} by a numerical 
procedure based on a shooting method,\cite{ES89} and more recently 
using the WKB approximation\cite{WKB97} and perturbation 
theory.\cite{BMBJAM98} 
As mentioned above,  for the  partition function calculation, 
{\it all} states are needed, bound and unbound, optically active 
and inactive. 
None of the above methods is suitable for analysis of shallow bound 
states and low-energy scattering states.

We use for calculation of the scattering phase shifts and bound state 
energies entering Eqs.~(\ref{eq03}) and (\ref{eq04}) the 2D formulation
\cite{PG97} of the variable phase method.\cite{Calogero} 
In this method the scattering phase shift and the function 
defining bound-state energies can be obtained as the large 
distance limit of the phase function, which satisfies the 
first-order, nonlinear Riccati equation originating 
from the radial Schr\"{o}dinger equation (see Appendix~\ref{VPM}).
The variable phase method is especially effective for calculation of 
the shallow-state binding energies and low-energy scattering phase 
shifts.

\begin{figure}
\begin{center}
\includegraphics{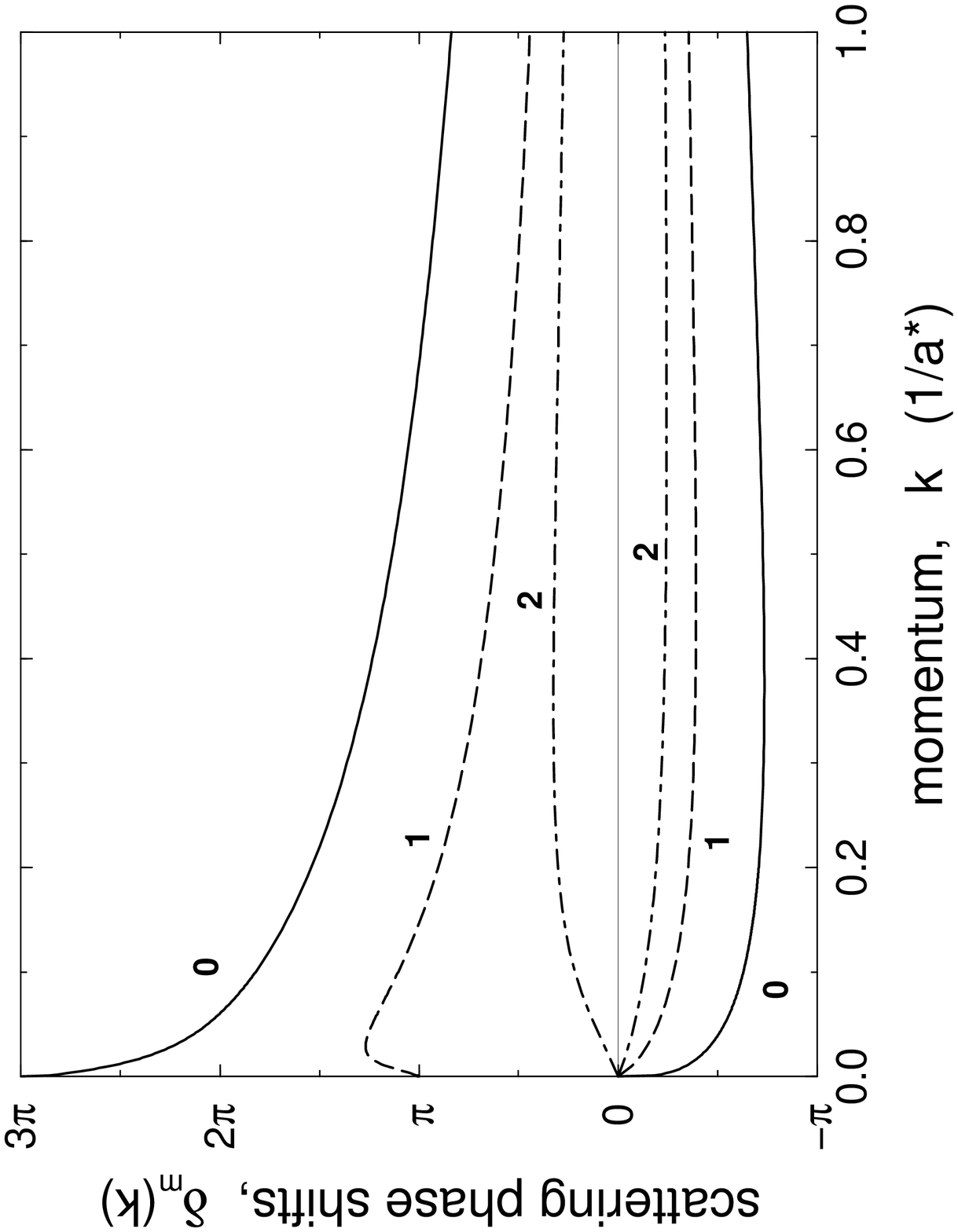}
\end{center}
\vskip 16truecm
\caption{
Scattering phase shifts versus the in-plane wave vector $~k~$ 
(in units of inverse Bohr radius $~1/a^*$) for a 2D particle in 
a screened Coulomb potential, Eq.(\ref{TF2D}).  
Screening wavenumber $~q_s=0.2/a^*$. 
For the attractive potential all phase shifts are positive and 
for the repulsive potential they are negative. Numbers show $~m~$ values.
}
\label{fig1}
\end{figure}

In Fig.~\ref{fig1} we show the $~k$-dependence of the scattering phase 
shifts for the attractive and repulsive Thomas-Fermi potentials [both 
signs of Eq.~(\ref{TF2D})] with the screening wavenumber $~q_s=0.2/a^*$,
where $~a^*=\epsilon \hbar^2/(\mu e^2)~$ is the 3D exciton Bohr radius.
The scattering phase shifts are negative for the repulsive potential and 
positive for the attractive potential. 
For the repulsive potential all zero-energy phase shifts vanish, 
$~\delta_m(k=0)=0~$ for all angular momenta $~m$.
For the attractive potential 
\begin{equation}  
\lim_{k \rightarrow 0}\delta_m(k)~=~\nu_m \pi~,
\label{Ltheor}
\end{equation}
where $~\nu_m~$ is the number of bound states.\cite{m1note1} 
Equation (\ref{Ltheor}) is the 2D analogue of Levinson's theorem 
\cite{Levinson,Newton} which connects the zero-energy scattering 
phase shift with the number of bound states for non-relativistic 
particles in 3D. 
This theorem has been known for almost five decades, 
however its applicability to the 2D scattering problem has been 
considered only recently.\cite{PG97,Bolle,Dong,Lin97}

We have recently \cite{PG97} used Levinson's theorem in the form of 
Eq.~(\ref{Ltheor}) for bound-state counting in the attractive Thomas-Fermi 
potential, Eq.~(\ref{TF2D}), and found a remarkably simple relation between 
the number of bound states and the screening wavenumber $~q_s$. 
With decreasing screening, new bound states appear at critical values 
of the screening length given by the simple formula \cite{PG97}
\begin{equation}
\left({1 \over q_s a^*}\right)_c~=~
{(2|m|+\nu-1)(2|m|+\nu) \over 2}~,~~~~\nu=1,2,~\ldots~, 
\label{qscrit}
\end{equation} 
Equation (\ref{qscrit}) can be easily inverted and the number of bound 
states for given $~m~$ and $~q_s~$ can be expressed as 
\begin{equation}
\nu_m~=~\max \left\{0~,~\nu_0~-2|m|\right\}~,
\label{numqs}
\end{equation} 
where
\begin{equation}
\nu_0~=~\Bigglb[{\sqrt{8/(q_s a^*)+1}~+~1 \over 2} \Biggrb]~,
\label{nu0}
\end{equation} 
is the number of bound states with $~m=0$.
Here, and in Eqs.~(\ref{mmax}) and (\ref{nb2}), the bold
square brackets designate the integer part of a number.
For small $~q_s$, Eq.~(\ref{nu0}) gives a  number 2.5 times smaller than 
the WKB estimate\cite{WKB97} for the maximum number of bound $s$-states.    
The Bargmann bound condition \cite{Bargmann} (re-stated for the 2D case 
\cite{SH67}) for the attractive potential (\ref{TF2D}) is 
$~\nu_m<1/(mq_sa^*)$. 
This was also found to give a gross over-estimate of the number of 
bound states. 

The total number of bound states, $~N_b$, for a given $~q_s a^*~$ 
\begin{equation}
N_b~=~\nu_0~+~2 \sum_{m=1}^{|m|_{max}}\nu_m~,
\label{nb1}
\end{equation}
can also be found explicitly as follows.
From Eq.~(\ref{qscrit}) the maximum possible value of $~|m|~$ 
for the state which remains bound is
\begin{equation}
|m|_{max}~=~\Bigglb[{\sqrt{8/(q_s a^*)+1}~-1 \over 4}\Biggrb]~=~
\bigglb[{\nu_0-1 \over 2}\biggrb]~.
\label{mmax}
\end{equation}
Then the sum in Eq.~(\ref{nb1}) can be easily evaluated 
using Eqs.~(\ref{numqs}) and (\ref{mmax}):
\begin{eqnarray}
N_b~&=&~\nu_0~+~2|m|_{max}(\nu_0-|m|_{max}-1)
\nonumber\\
~&=&~\nu_0~+~2 \bigglb[{\nu_0-1 \over 2}\biggrb] 
\left\{\nu_0-1-\bigglb[{\nu_0-1 \over 2}\biggrb]\right\}.
\label{nb2}
\end{eqnarray}
For small $~q_s a^*~$ (weakly screened potential) a simple 
approximate expression for the total number of bound states follows 
from substitution of Eq.~(\ref{nu0}) into Eq.~(\ref{nb2}):
\begin{equation}
N_b~\approx~{\nu_0^2 \over 2}~\approx~{1 \over q_s a^*}
\label{nb3} 
\end{equation}
Thus, for the weakly screened Thomas-Fermi potential the bound-state sum 
in the partition function, Eq.~(\ref{eq03}), has a finite number of terms 
which is approximately equal to the screening radius $~1/q_s~$ measured 
in units of the Bohr radius. 
The WKB estimate of the number of bound states\cite{WKB97} gives a 
different (square root) dependence of $~N_b~$ on $~1/(q_s a^*)~$ for 
small $~q_s a^*$. The reason for this difference is that in 
Ref.~\onlinecite{WKB97} only $~m=0~$ states are considered, whereas 
all values of $~m~$ are needed to obtain the result of Eq.~(\ref{nb3}).

\begin{figure}
\begin{center}
\includegraphics{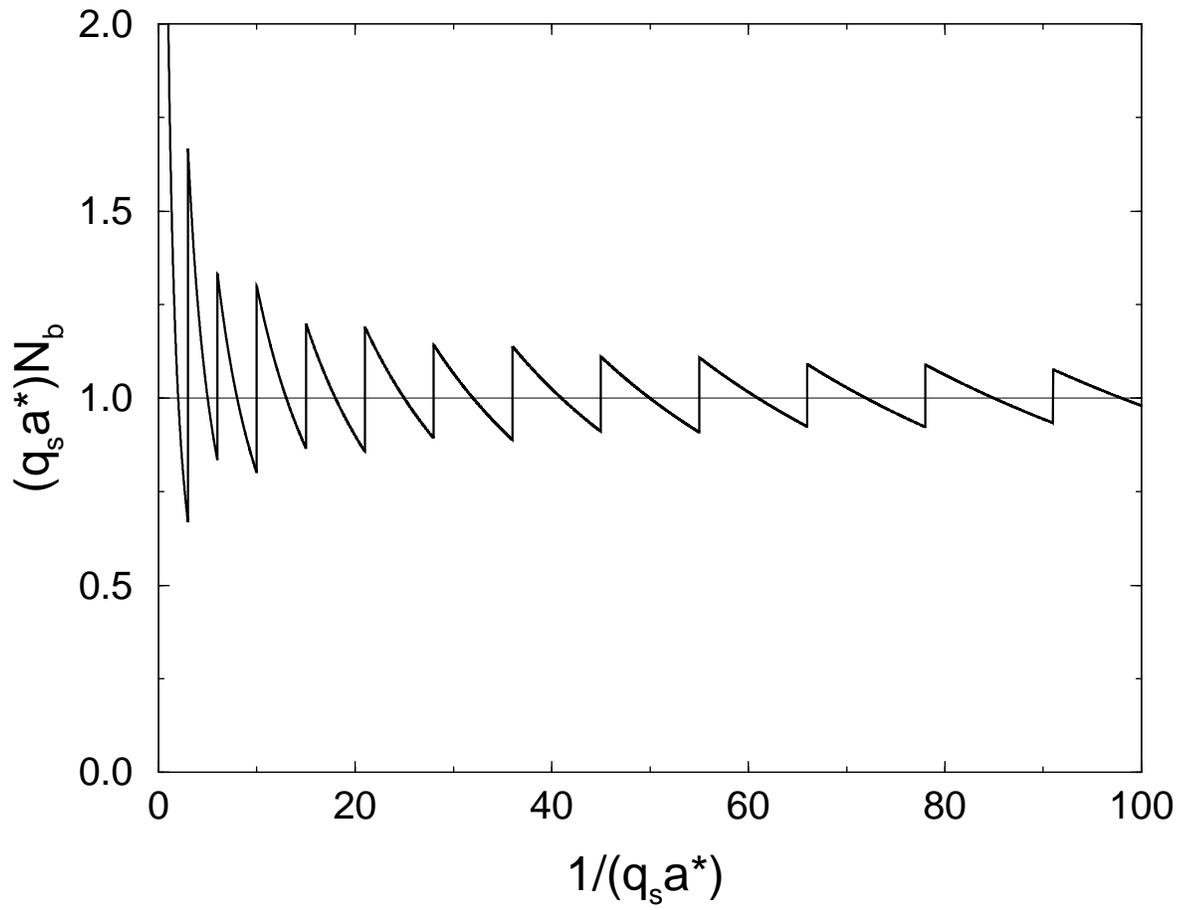}
\end{center} 
\vskip 16truecm
\caption{
The normalized number of bound states $~(q_s a^*) N_b~$ as a function 
of the inverse screening parameter, $~1/(q_s a^*)$, for $~q_sa^*\leq 1$.
}
\label{fig2}
\end{figure}

As the screening is reduced $~N_b~$ given by Eq.~(\ref{nb2}) exhibits 
steps of ever increasing height. 
In order for the limit of Eq.~(\ref{nb3}) to be meaningful the step 
height should be smaller than $~N_b~$ itself, i.e., the normalized number
of bound states, $~N_b/N_b^{q_s\rightarrow 0}=(q_s a^*)N_b$, 
should converge to unity as $~q_s a^*\to 0$. 
As can be seen in  Fig.~\ref{fig2} this number oscillates around 
unity with the amplitude of oscillations decreasing with increasing 
$~1/(q_s a^*)$. It can be shown that for $~q_s a^*\to 0~$ the 
amplitude of these oscillations is proportional to $~(q_sa^*)^{1/2}~$ 
and their period is proportional to  $~(q_sa^*)^{-1/2}$. 
 
\begin{figure}
\begin{center}
\includegraphics{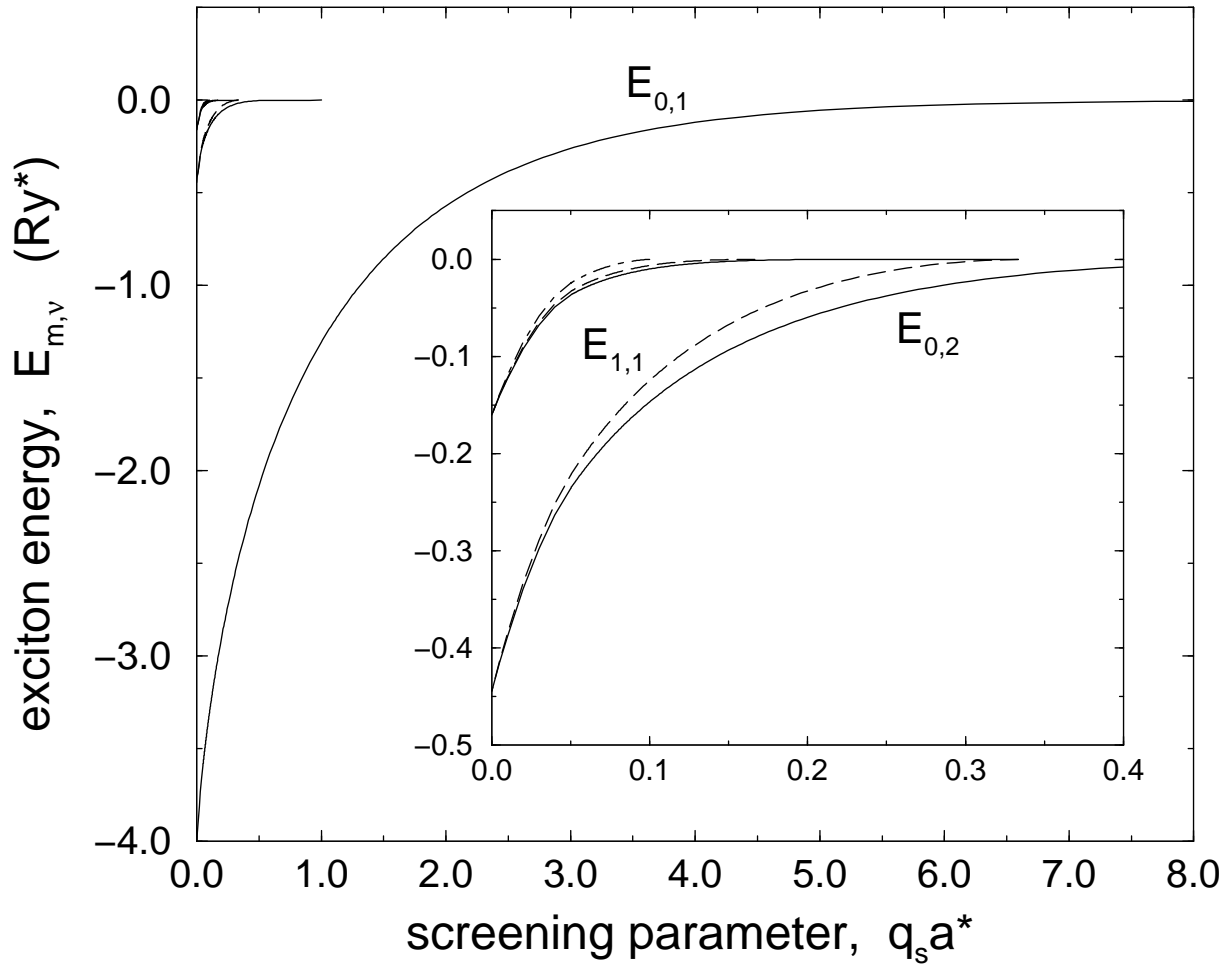}
\end{center} 
\vskip 16truecm
\caption{
The bound-state energies $~E_{m,\nu}~$ of the 2D exciton in exciton 
Rydberg units are shown as a function of the screening parameter 
$~q_sa^*~$ for different $~m~$ values. Solid lines show $~m=0~$ states
($E_{0,1}$, $~E_{0,2}$, and $~E_{0,3}$);
dashed lines show $~m=1~$ states ($E_{1,1}$ and $~E_{1,2}$);
the dot-dashed line shows the lowest state with $~m=2~$ ($E_{2,1}$).
Inset is a blowup near $~q_sa^*=0$.}
\label{fig3}
\end{figure}

In order to calculate the partition function, the bound state energies 
are required. 
These can also be obtained using the variable phase method, 
and the necessary equations are presented in Appendix~\ref{energy}.
Numerical results for the attractive screened Coulomb potential 
[upper sign in Eq.~(\ref{TF2D})] are presented in Fig.~\ref{fig3}. 
In this figure the energies $~E_{m,\nu}~$ of the several lowest bound 
states of the screened exciton are shown as a function of the screening 
wavenumber $~q_sa^*$. 
Here the energies are measured in effective exciton Rydberg 
[${\rm Ry}^*=\hbar^2/(2\mu_{eh}{a^*}^2)$] and we use the same 
classification of energy levels as in Ref.~\onlinecite{PG97}, 
i.e., each energy level is characterized by the angular momentum 
quantum number $~m~$ and another number $~\nu~$ which numerates 
different bound states for a given $~m$, with $~(\nu-1)~$ being 
the number of non-zero nodes of the radial wave function. 
For $~m=0~$ ($s$-states) the calculated energies are consistent with 
those obtained by J.~Lee {\it et al}\cite{LSpM85} using a variational 
method.
 
\section{PARTITION FUNCTIONS AND IONIZATION DEGREE}
\label{Results}

Before we present the results of calculations of the partition functions 
and the ionization degree, we would like to discuss an important 
consequence of Levinson's theorem for the statistical mechanics of the 
2D gas with an attractive interaction between its particles. 
The bound-state sum $~Z_{bound}=\sum_{m,\nu} \exp(-\beta E_{m,\nu})~$
entering the two-body partition function in Eq.~(\ref{eq03}) exhibits 
jumps whenever bound states disappear with increasing screening.
We will now show that these jumps do not give rise to unphysical 
discontinuities in the partition function if the scattering states 
are properly taken into account.
 
\begin{figure}
\begin{center}
\includegraphics{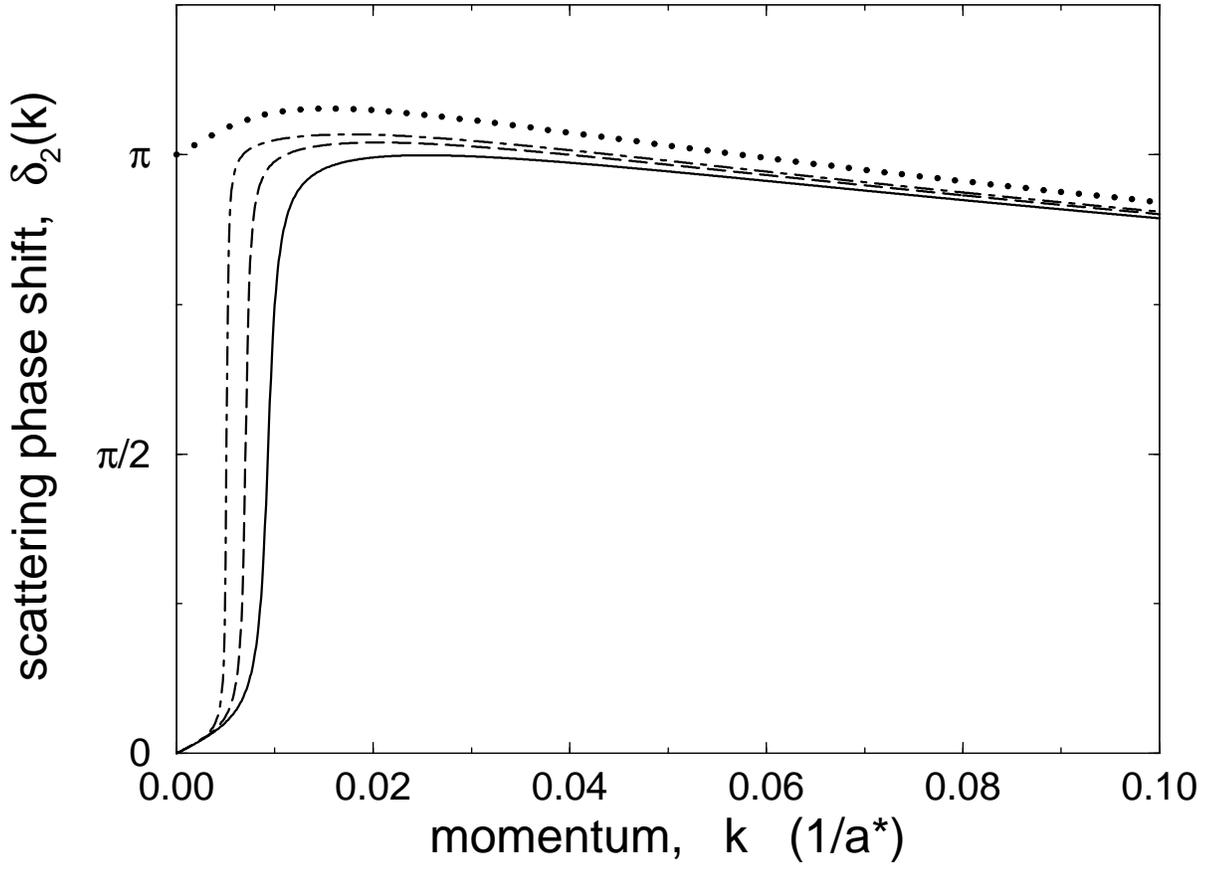}
\end{center} 
\vskip 16truecm
\caption{
The scattering phase shift $~\delta_2~$ is shown as a function of the 
in-plane wave vector $~k~$ (measured in inverse exciton Bohr radii)
for several values of the inverse screening parameter close to the 
critical value, $~1/(q_sa^*)=10$.
Solid line: $~1/(q_sa^*)=9.9$; dashed line: $~1/(q_sa^*)=9.95$; 
dot-dashed line $~1/(q_s a)=9.98$.
Dots show $~\delta_2(k)~$ for $~1/(q_sa^*)=10.1~$ (a shallow bound 
state with $~m=2~$ has just appeared).
}
\label{fig4}
\end{figure}

Integrating by parts the scattering term and using Levinson's theorem 
in the form of Eq.~(\ref{Ltheor}) we can rewrite Eq.~(\ref{eq03}) as 
\begin{eqnarray}
Z_{eh}&~=~&\sum_{m,\nu}\,\{\exp(-\beta E_{m,\nu})\,-\,1\} 
\nonumber\\
&~+~&{2 \over \pi q_T^2} \int_0^\infty \left(\sum_{m=-\infty}^{\infty} 
{\delta_m(k)} \right)\,\exp(-k^2 /q_T^2)\,kdk~,
\label{res1} 
\end {eqnarray}
where $~q_T^2=2\mu_{eh} k_B T/\hbar^2$.
The modified bound-state sum [the first term in Eq.~(\ref{res1})] does not 
exhibit jumps whenever bound states disappear with increasing screening. 
For non-zero temperature the scattering integral [the second term in 
Eq.~(\ref{res1})] is also a smooth function of the interaction strength, 
which can be understood from Fig.~\ref{fig4}. 
In this figure the scattering phase shift $~\delta_2~$ is plotted as a 
function of $~k~$ for several values of $~1/(q_sa^*)~$ close to the critical 
value $~1/(q_sa^*)=10~$ when the first bound state with $~m=2~$ appears. 
One can see that although $~\delta_m(0)~$ has a jump when $~q_s~$ passes 
a critical value this jump does not influence the value of the scattering 
integral if the thermal wavenumber $~q_T~$ is larger than the interval 
of $~k~$ in which $~\delta_m(k)~$ changes rapidly. 
As shown in Fig.~\ref{fig4}, when the bound state disappears the phase 
shift is affected only in an infinitesimally thin region around $k=0$. 
For any non-zero temperature this transition region makes no contribution 
to the phase-shift integral.
Thus the electron-hole interaction part of the partition function given 
by Eq.~(\ref{res1}) is a smooth function of the interaction strength as 
expected from the general thermodynamic argument.\cite{PGprb1,Gibson87}
Similar cancellation of the bound state sum discontinuities for a 
3D plasma is well known.\cite{Larkin60,Rogers71}  

\begin{figure}
\begin{center}
\includegraphics{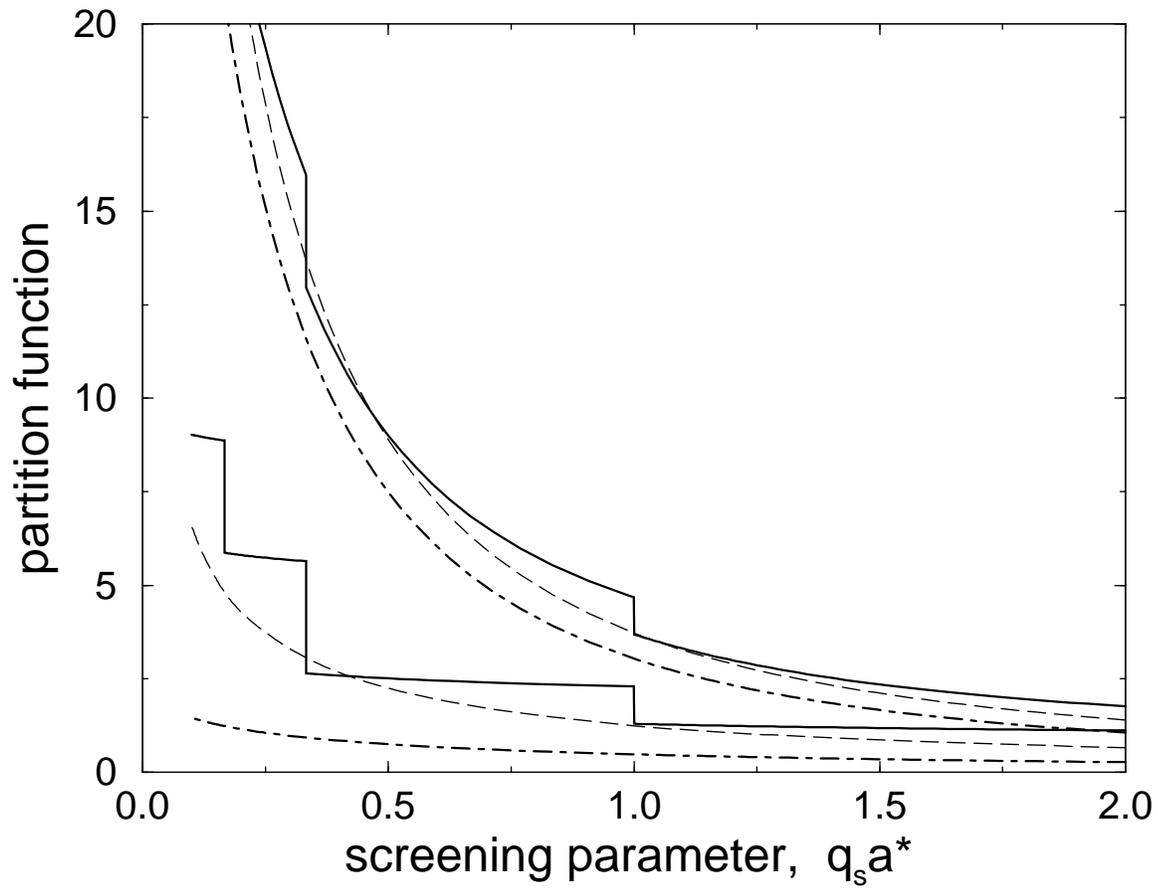}
\end{center} 
\vskip 16truecm
\caption{
The two-body interaction part of the partition function versus the 
inverse screening parameter $~1/(q_sa^*)~$ for two values of 
$~k_B T/{\rm Ry^*}$.
Three upper curves: $~k_B T=1~{\rm Ry}^*$; three lower curves: 
$~k_B T=5~{\rm Ry}^*$.
Solid lines show the bound state contributions $~Z_{bound}~$ only;
dashed lines show $~Z_{eh}$; dot-dashed lines show  $~Z_{eh}+Z_{ee}$.}
\label{fig5}
\end{figure}

The results of the calculation of the two-body interaction part of the 
partition function for the model semiconductor,\cite{Zimm,Littlewood3d}
for which the assumption $~M_e=M_h=2\mu_{eh}~$ is made, are presented in 
Fig.~\ref{fig5}. 
Calculations are performed for two values of the ratio of $~k_BT~$ to the 
bulk excitonic Rydberg, $~k_B T/{\rm Ry}^*=1~$ (three upper curves) and 
$~k_B T/{\rm Ry}^*=5~$ (three lower curves),  
which roughly correspond to ZnSe (or GaN) and GaAs at room temperature. 
Solid lines show the bound-state sum,     
$~Z_{bound}=\sum_{m,\nu} \exp(-E_{m,\nu}/k_B T)$, which exhibits jumps 
whenever bound states disappear with increasing screening.
The electron-hole part of the partition function, $~Z_{eh}$, which is 
shown by dashed lines, is a smooth function of the screening parameter, 
the bound state sum discontinuities are compensated by the scattering 
state contributions. 
Dot-dashed lines show the sum $~Z_{eh}+Z_{ee}$, which enters the 
modified law of mass action, Eq.~(\ref{eq02}) (when simplified for the 
model semiconductor). 
Note that the cancellation of the $~Z_{eh}~$ term by the $~Z_{ee}~$ term
for $~k_B T/{\rm Ry}^*=5~$ is stronger than for $~k_B T/{\rm Ry}^*=1$. 
This can be explained by the enhanced role of scattering states 
for the higher ratio of $~k_BT~$ to the excitonic Rydberg.
The lower absolute value of $~Z_{eh}+Z_{ee}~$ ensures that thermodynamic 
properties of the 2D electron-hole plasma in GaAs are much closer to the 
ideal gas behavior than those in the case of the wide-gap semiconductor.
\cite{PGprb1}   

For the two-component electron-hole plasma the Thomas-Fermi 2D screening 
wavenumber entering Eq.~(\ref{TF2D}) is given  
in the  Boltzmann limit by \cite{LSpM85}
\begin{equation}
q_sa^*~=~
{2\pi\hbar^2 \over \mu_{eh} k_B T}\,
(n_h^0+n_e^0)
~=~4\pi\,{{\rm Ry}^*\over k_B T}\,
\left(n_e^0{a^*}^2+n_h^0{a^*}^2\right).
\label{qsboltzm}
\end{equation}
Note that we use uncorrelated quasiparticle densities $~n_e^0~$ and
$~n_h^0~$ for the calculation of the screening wavenumber, since 
Eq.~(\ref{qsboltzm}) is derived for the non-interacting 2D plasma. 
\cite{Stern67,HKbook}
We assume that the screening by excitons is much smaller then the free 
carrier screening when exciton and free carrier densities are of the 
same order.
However, if one calculates the screening wavenumber using the difference 
between the total number of carriers and the number of bound carriers, 
unphysical jumps appear in the dependence of the screening wavenumber 
on total density as shallow bound states disappear with increasing density. 
Thus, it is natural to calculate $~q_s~$ on the basis of 
uncorrelated density ($n_e^0+n_h^0$), which is a part of the 
total density behaving as an ideal gas (see Appendix~\ref{BUformula}) 
and which is a smooth function of the total density.
 
\begin{figure}
\begin{center}
\includegraphics{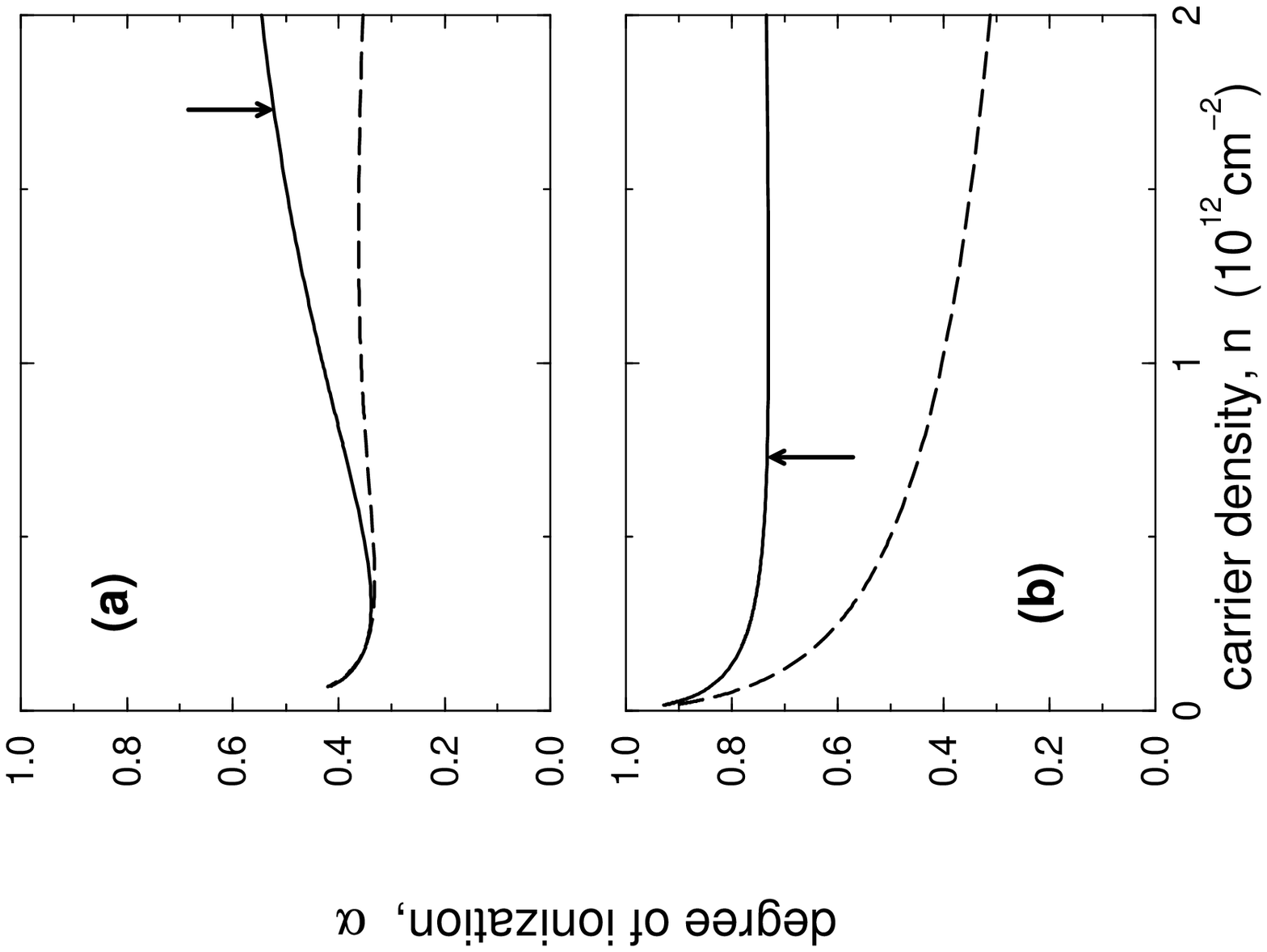}
\end{center}
\vskip 16truecm
\caption{The degree of ionization (solid lines) of the nondegenerate 
2D electron-hole plasma as a function of the total electron density
at room temperature, calculated for the model semiconductor with 
the effective Bohr radius and excitonic Rydberg of (a) ZnSe, and (b) GaAs. 
The arrows indicate $~n=2/\lambda^2_{M_e}~$ for ZnSe and GaAs at room 
temperature.
Dashed lines show the degree of ionization calculated using a simple
law of mass action with a single bound state.
}
\label{fig6}
\end{figure}

For the model semiconductor, $~n_h^0=n_e^0=\alpha\,n$, and
Eq.~(\ref{qsboltzm}) can be further simplified to
\begin{equation}
q_sa^*~=~8\pi\,\alpha\,{{\rm Ry}^*\over k_BT}\,n{a^*}^2~.
\label{qsmodel}
\end{equation}
Equation (\ref{qsmodel}) shows clearly the connection between the 
dimensionless screening parameter $~q_sa^*~$ and the two main dimensionless 
parameters characterizing the 2D electron-hole plasma, namely the 
dimensionless density $~n{a^*}^2~$ and temperature $~k_B T/{\rm Ry}^*$. 
In addition the role of the degree of ionization, $~\alpha~$, introduced 
by Eq.~(\ref{alpha0}) becomes more transparent. 
The parameter $~\alpha~$ enters Eq.~(\ref{qsmodel}) explicitly, 
governing the screening wavenumber which determines the strength 
of the interaction between charged particles in the plasma. 
In turn the degree of ionization itself depends on $~q_sa^*~$ through 
the partition functions $~Z_{eh}~$ and $~Z_{ee}$.  
For the model semiconductor the modified law of mass action, 
Eq.~(\ref{eq02}), can be rewritten as
\begin{equation}  
n_e^{corr}{a^*}^2 ~=~4\pi\,
{\left(n_e^0{a^*}^2\right)}^2\,{{\rm Ry}^*\over k_BT}\,
\left(Z_{eh}+Z_{ee}\right), 
\label{modsemlaw}
\end{equation}
and using Eqs.~(\ref{alpha0}) and (\ref{qsmodel}) we get the following 
expression for the degree of ionization $~\alpha$:
\begin{equation}
\alpha~=~\left\{1+{q_sa^*\over 2}\,
(Z_{eh}+Z_{ee})\right\}^{-1}~.
\label{modsemalpha}
\end{equation}
Equation (\ref{modsemalpha}) together with Eqs.~(\ref{eq03}) and 
(\ref{eq04}) for the partition functions allows us to calculate the 
degree of ionization $~\alpha~$ of the dilute (nondegenerate) 2D 
electron-hole plasma as a function of the screening parameter $~q_sa^*$. 
The connection between $~q_sa^*~$ and the total electron (hole) density 
$~n~$ [Eq.~(\ref{qsmodel})] can be used for self-consistent calculations 
of $~\alpha~$ as functions of $~n~$ for different temperatures or 
material parameters. 
The result of these calculations are shown in Fig.~\ref{fig6}. 
Calculations are performed for the model semiconductor with the exciton 
Bohr radius $~a^*~$ and effective Rydberg $~{\rm Ry}^*~$ corresponding 
to ZnSe (Fig.~\ref{fig6}a) and GaAs (Fig.~\ref{fig6}b) and for 
room temperature ($k_BT=300~$K). 
The arrows indicate the points of crossover from Boltzmann  
to Fermi statistics, $~n\,=\,2/\lambda^2_{M_e}$. 
As mentioned in Section~\ref{intro}, the nondegenerate treatment 
is more adequate for the wide-gap material.  

On the same plot we show by the dashed lines the degree of ionization 
calculated using a simple law of mass action with a single bound state 
(the ground state of the screened exciton). 
It can be seen from the figure that the degree of ionization is well 
described by the single-bound-state mass action law only for low carrier 
densities.\cite{Surprise}
For high densities (but remaining in the nondegenerate regime) the role of
scattering states becomes essential. 
Instead of the unphysical behavior predicted by the simple mass action law, 
in which the degree of ionization decreases with increasing density,
we find that the degree of ionization increases at higher densities. 

A minimum on the curve showing the density dependence of the degree of 
ionization has the following explanation. 
At low densities the main contribution to the correlated density comes from 
the ground exciton state,  which is almost unscreened. 
This state in 2D is at least nine times deeper than the first excited state. 
Therefore, the simple single-bound-state law of mass action is a good 
approximation at low densities, but not as $~n\,\rightarrow\,0~$ -- when 
the number of bound states becomes larger than the ground state contribution 
to the partition function, see Ref.~\onlinecite{Surprise}. 
The standard law of mass action states that the density of bound states 
is proportional to uncorrelated density squared, which reflects the fact 
that at fixed temperature (room temperature in our case) and at low density 
most of carriers occupy the high-energy ionized states in the continuum 
rather than the bound states. 
The low-density high-temperature electron-hole plasma behaves as an ideal 
gas with the degree of ionization close to unity. 
Thus, at low density the correlated density is proportional to the square 
of the total density and the degree of ionization decreases with increasing 
density. 
However, with the further increase in the total density screening becomes 
important and the inter-particle correlation caused by the Coulomb 
interaction starts to decrease. 
Correspondingly, the degree of ionization changes the character of its 
density dependence. 
There is a certain value of density, which corresponds to the 
minimal value of the degree of ionization.

As expected, in wide-gap semiconductors the calculated degree of 
ionization is much lower than in GaAs for the same temperature and 
carrier density. 
For both materials the calculated degree of ionization of the 
room-temperature 2D electron-hole plasma reaches its minimum 
at a certain density. 
The same happens for a 3D plasma,\cite{Zimm,PGcrgr} however the 
minimal value of the degree of ionization for the 3D plasma is 
much higher than in the 2D case (compare Fig.~\ref{fig6} with
Fig.~1 in Ref.~\onlinecite{PGcrgr}). 
This is due to the much enhanced binding energy in 2D.

The inclusion of Fermi statistics and phase-space filling, which 
is beyond the scope of the present paper, would provide a sharper 
rise of $~\alpha~$ at high carrier densities as the phase space 
available for the construction of exciton states is restricted. 
This will apply to both wide-gap and narrow-gap semiconductors. 
In the foregoing discussion we have assumed a purely 2D plasma. 
This assumption gives an overall overestimate of exciton binding 
energies, compared to a real, finite-width quantum well, for which 
unscreened exciton binding energies are lower and the finite 
thickness correction enhances the screening effect.\cite{Henriques}
Thus, the results shown in Fig.~\ref{fig6} should be considered as 
lower bound estimates of the degree of ionization of the electron-hole 
plasmas in ZnSe and GaAs quantum wells at room temperature. 
Note, that even this lower estimate does not give a value of the 
degree of ionization of a plasma in a ZnSe quantum well below 0.33.
It means that at room temperature at least one third of the carriers are
always unbound, which has to be taken into account in gain calculations. 

\section{CONCLUSION}
\label{Conclusion}

We have calculated the degree of ionization of the 2D electron-hole 
plasma taking into account {\it all} screened exciton bound states 
as well as scattering states.
It has been shown that the scattering state contribution changes the 
character of the density dependence of the degree of ionization.
We have found that the degree of ionization of the 2D plasma reaches 
its minimal value at intermediate densities and approaches unity at 
high densities, which differs from the result based on the simple 
law of mass action.  

The calculated degree of ionization of the electron-hole plasma in 
a ZnSe quantum well is significantly lower than in a GaAs quantum 
well with the same carrier density and temperature. 
Therefore, excitonic processes should be considered for gain 
calculations in quantum wells based on wide-gap semiconductors. 
However, at room temperature at least one third of the carriers in ZnSe 
wells is shown to be unbound, which allows us to speculate that the most 
likely lasing mechanism at moderate density is exciton/free-carrier 
scattering.

Most of the results presented here are obtained for the model system 
with equal electron and hole effective masses. 
For wide-gap semiconductors at room temperature ($k_B T\sim1~{\rm Ry}^*$) 
this approximation is good, since $~Z_{ee}~$ is much smaller than  
$~Z_{eh}~$ and the influence of the electron-electron part of the 
partition function on the degree of ionization is not significant. 
In the case of an extreme difference between electron and hole 
masses the model fails, e.g., lighter quasiparticles can be 
degenerate, when heavy quasiparticles are nondegenerate. 

The variable phase method is a powerful tool for studying scattering 
and bound states in any short-range potential. 
This method enabled us to find hitherto undiscovered properties of 
a Coulomb potential statically screened by a 2D electron gas. 
The same approach can be applied to a more refined potential, 
which takes into account Friedel oscillations and the finite 
thickness of the 2D layer.
   
\acknowledgments

This work was supported by the UK EPSRC and the Royal Society,
and we thank Dr. S.-C. Lee and H. Ouerdane for a careful reading 
of the manuscript.

\appendix
\section{VARIABLE PHASE METHOD IN TWO DIMENSIONS}
\label{VPM} 

In this Appendix we derive the basic equations of the variable phase 
approach in two dimensions from the radial Schr\"{o}dinger equation. 
This derivation is similar to that in 3D.\cite{Calogero}

\subsection{Scattering phase shifts}
\label{shifts} 

The relative in-plane motion of two interacting particles with masses 
$~M_a~$ and $~M_b~$ and the energy of relative motion $~E~$ can be 
considered as a motion of a particle with the mass 
$~\mu_{ab}=M_a M_b/(M_a+M_b)~$ and energy $~E$, moving in an external 
central potential $~V(\rho)$. 
This motion is described by the wave function satisfying the stationary 
Schr\"{o}dinger equation  
\begin{equation}                
\hat{H}_{rel}\psi=-{\hbar^2\over 2\mu_{ab}}\left({1\over\rho}
{\partial\over\partial\rho}\rho{\partial\over\partial\rho}
+{1\over\rho^2}{\partial^2\over\partial\varphi^2}
\right)\psi+V(\rho)\psi=E\psi~.
\label{vpm1}
\end{equation}
Owing to the axial symmetry of the potential $~V(\rho)$, we can separate 
variables in the expression for the wave function
\begin{equation}
\psi_m(\rho,\varphi)=R_m(\rho)e^{i m\varphi},~~~~~m=0,\pm1,\pm2,~\ldots~.
\label{vpm2}
\end{equation}
The  equation for the radial function $~R_m(\rho)~$ reads
\begin{equation}
R''_m+{1\over\rho}R'_m+
\left(k^2-U(\rho)-{m^2\over\rho^2}\right)R_m~=~0~,
\label{vpm3}
\end{equation}
where $~k^2=2\mu_{ab}E/\hbar^2~$ and $~U(\rho)=2\mu_{ab}V(\rho)/\hbar^2$.
In what follows we consider $m\geq0$ only, 
as $~R_{-m}(\rho)=R_{m}(\rho)$.

We assume that the interaction potential vanishes at infinity
(the precise decay rate will be discussed later).
Then at large distances the radial function satisfies the free 
Bessel equation, whose general solution is 
\begin {eqnarray} 
R_m(\rho)&&~=~A_m[J_m(k\rho) \cos{\delta_m}
-N_m(k\rho) \sin{\delta_m}]~ 
\nonumber\\ 
&&\stackrel{\rho\rightarrow\infty}{\longrightarrow}~ 
A_m\left({2\over\pi k \rho}\right)^{1/2} 
\cos(k\rho - (2m+1)\pi/4 + \delta_m)~,~ 
\label{vpm4} 
\end{eqnarray} 
where $~\delta_m~$ is the scattering phase shift,\cite{SH67,LandauQM}
$~J_m(k\rho)~$ and $~N_m(k\rho)~$ are the Bessel and the Neumann 
functions, respectively.

In the variable phase approach, $~A_m~$ and $~\delta_m~$ are considered 
not as constants but as functions of the distance $~\rho$. 
The amplitude function $~A_m(\rho)~$ and the phase function 
$~\delta_m(\rho)~$ are introduced by the equation
\begin{equation}
R_m(\rho)~=~A_m(\rho)[J_m(k\rho)\cos{\delta_m(\rho)}
-N_m(k\rho)\sin{\delta_m(\rho)}]~, 
\label{vpm5}
\end{equation}
with the additional condition, which we are free to choose as
\begin {equation}
R'_m(\rho)~=~A_m(\rho)[J'_m(k\rho)\cos{\delta_m(\rho)} 
-N'_m(k\rho)\sin{\delta_m(\rho)}]~,
\label{vpm6} 
\end{equation}
where the prime indicates differentiation with respect to $~\rho$.
The phase function $~\delta_m(\rho)~$ has a natural physical 
interpretation as being the phase shift produced by a potential 
cut-off at a distance $~\rho$.

Differentiating Eq.~(\ref{vpm6}) and substituting the resulting 
expression, together with Eqs.~(\ref{vpm5}) and 
(\ref{vpm6}), into Eq.~(\ref{vpm3}) we get
\begin{eqnarray}
A'_m(\rho)[J'_m(k\rho)\cos{\delta_m(\rho)}
-N'_m(k\rho)\sin{\delta_m(\rho)}]~~~~~~~~~~~~~~~~~~~~~~~~~
\nonumber\\
-\delta'_m(\rho)A_m(\rho)[J'_m(k\rho)\sin{\delta_m(\rho)} 
+N'_m(k\rho)\cos{\delta_m(\rho)}]~~~~~~~~~~~~~
\nonumber\\
=U(\rho)A_m(\rho)[J_m(k\rho)\cos{\delta_m(\rho)}
-N_m(k\rho)\sin{\delta_m(\rho)}]~.
\label{vpm7} 
\end{eqnarray}
To obtain Eq.~(\ref{vpm7}) we used that the functions $~J_m(k\rho)~$
and $~N_m(k\rho)~$ satisfy the free Bessel equation:
\[
F''_m+{1\over\rho}F'_m+\left(k^2-{m^2\over\rho^2}\right)F_m
~=~0~.
\]

Equating the derivative of Eq.~(\ref{vpm5}) to Eq.~(\ref{vpm6}) 
implies the following condition on the derivatives 
of the amplitude and the phase functions:
\begin{eqnarray}
A'_m(\rho)[J_m(k\rho)\cos{\delta_m(\rho)}
-N_m(k\rho)\sin{\delta_m(\rho)}]~~~~~~~~~~~~~~~~~~~~~~~~~
\nonumber\\
=\delta'_m(\rho)A_m(\rho)[J_m(k\rho)\sin{\delta_m(\rho)}
+N_m(k\rho)\cos{\delta_m(\rho)}]~.
\label{vpm8} 
\end{eqnarray}
Substituting $~A'_m(\rho)$, obtained from Eq.~(\ref{vpm8}), in 
Eq.~(\ref{vpm7}) yields
\begin{eqnarray}
-\delta'_{m}(\rho)[J_m(k\rho)N'_m(k\rho)-N_m(k\rho)J'_m(k\rho)]
~~~~~~~~~~~~~~~~~~~~~~~~~~~~~
\nonumber\\
=U(\rho)[J_m(k\rho)\cos{\delta_m(\rho)}
-N_m(k\rho)\sin{\delta_m(\rho)}]^2~.
\label{vpm9} 
\end{eqnarray}
Equation (\ref{vpm9}) can be simplified further, using the Wronskian 
of the Bessel functions
\[
W\{J_m(x), N_m(x)\}~=~J_m(x)\,{d\over dx}N_m(x)\,-
\,N_m(x)\,{d\over dx}J_m(x)~=~{2\over\pi x}~,
\]
and thus  becomes
\begin{eqnarray}
{d\over d\rho}\,
\delta_m(\rho)~&=&~-{\pi\over 2}\,\rho\,U(\rho)
\nonumber\\
&\times&~[J_m(k\rho)\cos{\delta_m(\rho)}
-N_m(k\rho)\sin{\delta_m(\rho)}]^2~.
\label{vpm10} 
\end{eqnarray}
This {\it phase equation}, Eq.~(\ref{vpm10}), is a first-order,
non-linear differential equation of the Ricatti type, which must 
be solved with the initial condition 
\begin {equation} 
\delta_m(0)~=~0~, 
\label{vpm11} 
\end{equation} 
thus  ensuring  that the radial function does not diverge at 
$~\rho=0$.
The total scattering phase shift $~\delta_m~$ can be obtained as a 
large distance limit of the phase function $~\delta_m(\rho)$:
\begin {equation}
\delta_m~=~\lim_{\rho \rightarrow \infty}\delta_m(\rho)~.
\label{vpm12}
\end {equation}
For numerical convenience, instead of the initial condition 
Eq.~(\ref{vpm11}), the small-$\rho~$ expansion is used 
\begin {equation} 
\delta_m(\rho)~\approx~-{\pi k^{2m} \over 2^{2m+1} (m!)^2} 
{\int}_{0}^{\rho}U(\rho^\prime){\rho^\prime}^{2m+1}d\rho^\prime~, 
~~~\rho~\rightarrow~0~. 
\label{vpm13}
\end {equation}

From Eq.~(\ref{vpm10}) and the asymptotic expansions of the Bessel 
functions one can see that the variable phase method is applicable 
only if the scattering potential $~U(\rho)~$ satisfies the 
necessary conditions:
\begin{equation}
{\int}_{\rho}^{\infty}U(\rho^\prime)~d\rho^\prime~\rightarrow~0~, 
~~~\rho~\rightarrow~\infty~,~ 
\label{vpm14} 
\end{equation} 
and 
\begin {equation} 
{\rho}^{2}U(\rho)~\rightarrow~0~, ~~~\rho~\rightarrow~0~.~ 
\label{vpm15} 
\end {equation}
The statically screened Coulomb potential $~V_s(\rho)$, defined by 
Eq.~(\ref{TF2D}), behaves like $\rho^{-1}$ at small distances 
and like $\rho^{-3}$ at large distances. Such behavior allows 
the application of the variable phase method to this potential.

\subsection{Bound state energies}
\label{energy}

For the states with negative energy of the relative motion (bound states),
the wavenumber $~k~$ is imaginary, $~k~=~i\kappa$,        
and we introduce the function $~\eta_m (\rho,\kappa)~$ 
vanishing in the origin and satisfying a non-linear equation  
\begin{eqnarray}
{d \over d\rho}\,\eta_m(\rho,\kappa)~&=&~-{\pi \over 2}\,\rho\,U(\rho)
\nonumber\\
&\times&~\left[ I_m(\kappa \rho)\cos{\eta_m(\rho,\kappa)}+ 
{2 \over \pi} K_m(\kappa \rho)\sin{\eta_m(\rho,\kappa)}\right]^2~,~~ 
\label{vpmb1}
\end{eqnarray}
where $~I_m(\kappa \rho)~$ and $~K_m(\kappa \rho)~$ are the modified Bessel 
functions of the first and  second kind, respectively. 
Equation (\ref{vpmb1}) is derived in the same fashion as 
Eq.~(\ref{vpm10}).
The functions $~I_m(\kappa \rho)~$ and $~K_m(\kappa \rho)~$ represent two 
linearly independent solutions of the free radial-wave Schr\"{o}dinger 
equation for the negative value of energy, $~E=-\hbar^2\kappa^2/2\mu_{ab}$, 
and $~\cot{\eta_m}~$ characterizes the weights of the diverging 
[$I_m(\kappa\rho)$] and converging [$K_m(\kappa\rho)$] 
solutions as $~\rho \rightarrow \infty~$. 
For the bound state, the diverging solution vanishes, 
implying the asymptotic condition 
\begin {equation}
\eta_m(\rho \rightarrow \infty,\,\kappa_\nu)~=~(\nu-1/2)\,\pi~,
~~~~\nu=1,2,~\ldots~.
\label{vpmb2} 
\end {equation} 
Here $\nu$ numerates the bound states for a given $m$ and $~(\nu-1)~$ is the 
number of non-zero nodes of the radial wave function. For numerical solution 
of Eq.~(\ref{vpmb1}), instead of the boundary condition $~\eta_m(0,\kappa)=0$, 
an asymptotic initial condition [analogous to the condition Eq.~(\ref{vpm13}) 
for the phase function $~\delta_m(\rho)$] is used.         

\section{BETH-UHLENBECK FORMULA IN TWO DIMENSIONS}
\label{BUformula}

In this Appendix we derive Eq.~(\ref{eq03}), which is the 2D analogue 
of the Beth-Uhlenbeck formula,\cite{BU37} and the modified law of 
the mass action, Eq.~(\ref{eq02}). 
This derivation is similar to the analysis used for the calculation 
of the second virial coefficient of low-density $~^3$He and $~^4$He 
monolayers on graphite.\cite{SS73D75} 

Let us consider a binary mixture of components $~a~$ and $~b~$
in two dimensions. 
The grand partition function of the mixture is given by
\begin{equation}
\Omega(z_a,z_b,A,T)~=~\sum_{N_a,\,N_b}\,{\cal Q}_{N_a,N_b}\,
z_a^{N_a}\,z_b^{N_b}~,
\label{grandpf}
\end{equation}
where $~z_a~$ and $~z_b~$ are the fugacities ($z_a=e^{\beta\mu_a}$, 
with $~\mu_a~$ being the chemical potential of the component $~a$),
$~A~$ is the area of the 2D system, and $~{\cal Q}_{N_a,N_b}~$ is 
the partition function defined as
\begin{equation}
{\cal Q}_{N_a,N_b}(A,T)~=~{\rm Tr}\,e^{-\beta\hat{H}(N_a,N_b)}~,
\label{pfNaNb}
\end{equation}
where the trace is to be taken over all states of the system that has 
$~N_a~$ particles of the type $~a~$ and $~N_b~$ particles of the type 
$~b~$ in the area $~A$.

We now expand the quantity $~\ln{\Omega}~$ as a power series 
in $~z_a$, $~z_b$
\begin{equation}
\ln{\Omega(z_a,z_b,A,T)}~=~\sum_{l_a,\,l_b}\,A\,{\cal C}_{l_a,l_b}\,
z_a^{l_a}\, z_b^{l_b}~,
\label{loggrandpf}
\end{equation} 
The density of the component $~a~$ is given by 
\begin{equation}
n_a~=~{1\over A}\,z_a\,{d\over dz_a}\,\ln{\Omega}~=~
\sum_{l_a,l_b}\,l_a{\cal C}_{l_a,l_b}\,z_a^{l_a}\,z_b^{l_b}~.
\label{naexact}
\end{equation}
From this point we consider the low-density limit,
$~z_a,~z_b~\ll~1$, and neglect all the terms higher than $~z^2~$ in 
Eq.~(\ref{naexact}). Then
\begin{equation}
n_a~\approx~{\cal C}_{1,0}\,z_a\,
+\,{\cal C}_{1,1}\,z_a\,z_b\,+\,2\,{\cal C}_{2,0}\,z_a^2~. 
\label{naapprox}
\end{equation} 
From comparing corresponding powers in Eqs.~(\ref{loggrandpf}) and 
(\ref{grandpf}) we get 
\begin{equation}
{\cal C}_{1,0}~=~{\cal Q}_{1,0}/A~,
\label{c10_1}
\end{equation}
\begin{equation}
{\cal C}_{1,1}~=~\left({\cal Q}_{1,1}-
{\cal Q}_{1,0}{\cal Q}_{0,1}\right)/A~,
\label{c11_1}
\end{equation}
and
\begin{equation}
{\cal C}_{2,0}~=~\left({\cal Q}_{2,0}-
{1\over 2}{\cal Q}^2_{1,0}\right)/A~.
\label{c20_1}
\end{equation}
The next step is to calculate the partition functions entering 
Eqs.~(\ref{c10_1})-(\ref{c20_1}).
First of all the one-particle partition function, $~{\cal Q}_{1,0}$, is 
given by
\begin{equation}
{\cal Q}_{1,0}(A,T)~=~g_a\,{A\over(2\pi)^2}\,
\int{\,d^2{\bf k}\,\exp\left(-\beta{\hbar^2 k^2 \over 2M_a}\right)}
~=~g_a\,{A\over\lambda_{M_a}^2}~,
\label{Q10}
\end{equation}
where $~g_a~$ is a quantum state degeneracy and $~\lambda_{M_a}~$ is 
a thermal wavelength,
\begin{equation}
\lambda_{M_a}^2~=~{2\pi\beta\hbar^2 \over M_a}~.
\label{lambdaMa}
\end{equation}
This yields
\begin{equation}
{\cal C}_{1,0}~=~{g_a \over \lambda_{M_a}^2}~.
\label{c10_2}
\end{equation}

In order to find the two-particle partition function, $~{\cal Q}_{1,1}$,
it is useful to separate the center of mass motion and the relative 
motion of the two particles:
\begin{eqnarray}
{\cal Q}_{1,1}(A,T)~&=&~g_a\,g_b\,{A\over\lambda_{M_a+M_b}^2}\,
{\rm Tr}e^{-\beta\hat{H}_{rel}}
\nonumber\\
&=&~g_a\,g_b\,{A\over\lambda_{M_a+M_b}^2}\,
\int d^2 \bbox{\rho} \sum_n|\psi_n(\bbox{\rho})|^2\,e^{-\beta E_n}~,
\label{Q11int}
\end{eqnarray}
where the factor $~A/\lambda_{M_a+M_b}^2~$ appears from
performing the summation over all center-of-mass momenta, 
the Hamiltonian $~\hat{H}_{rel}~$ of the relative motion is 
given in Appendix~\ref{VPM}, and the sum in Eq.~(\ref{Q11int}) 
is taken over all different solutions of Eq.~(\ref{vpm1}). 

For the corresponding two-body system of noninteracting
distinguishable particles, one would have
\begin{equation}
{\cal Q}^{(0)}_{1,1}(A,T)~=~
g_a\,g_b\,{A\over\lambda_{M_a+M_b}^2}\,
\int d^2 \bbox{\rho} \sum_n|\psi^{(0)}_n(\bbox{\rho})|^2\,
e^{-\beta E^{(0)}_n}~,
\label{Q11nonint}
\end{equation}
where the superscript $~^{(0)}~$ refers to quantities of the 
noninteracting system.
The two-body {\it interaction} part of the partition function
is then defined by
\begin{eqnarray}
Z_{ab}~&=&~\int d^2 \bbox{\rho} \sum_n\left\{
|\psi_n(\bbox{\rho})|^2\,e^{-\beta E_n}\,-\,
|\psi^{(0)}_n(\bbox{\rho})|^2\,e^{-\beta E^{(0)}_n}\right\}
\nonumber\\
&=&~\sum_n\left\{e^{-\beta E_n}\,-\,e^{-\beta E^{(0)}_n}\right\}~.
\label{defZab}
\end{eqnarray}
Thus
\begin{equation}
{\cal Q}_{1,1}(A,T)~=~{\cal Q}^{(0)}_{1,1}(A,T)\,+\,
g_a\,g_b\,{A\over\lambda_{M_a+M_b}^2}\,Z_{ab}~.
\label{Q11viaZab}
\end{equation}

To analyze Eq.~(\ref{defZab}) further we must study the energy 
spectra $~E^{(0)}_n~$ and $~E_n$. For the noninteracting system, 
$~E^{(0)}_n~$ forms a continuum. We write 
\begin{equation}
E^{(0)}_n~=~{\hbar^2 k^2 \over 2\mu_{ab}}~,
\label{e0n}
\end{equation}
which defines the relative wave number $~k$. Then for the system 
of two noninteracting distinguishable particles the 
function $~{\cal Q}^{(0)}_{1,1}~$ given by Eq.~(\ref{Q11nonint}) 
can be easily evaluated as 
\begin{eqnarray}
{\cal Q}^{(0)}_{1,1}(A,T)~&=&~g_a\,g_b\,{A\over\lambda_{M_a+M_b}^2}\,
{A\over(2\pi)^2}\,\int{\,d^2{\bf k}\,
\exp\left(-\beta{\hbar^2 k^2 \over 2\mu_{ab}}\right)}
\nonumber\\
&=&~g_a\,g_b\,{A\over\lambda_{M_a+M_b}^2}\,
{A\over\lambda_{\mu_{ab}}^2}
~=~\left(g_a\,{A\over\lambda_{M_a}^2}\right)\,
\left(g_b\,{A\over\lambda_{M_b}^2}\right)
~=~Q_{1,0}\,Q_{0,1}~.
\label{Q11eval}
\end{eqnarray}
For the interacting system, the spectrum of $~E_n~$ in general 
contains a discrete set of values $~E_B$, corresponding to 
two-body bound states, and a continuum. In the continuum, we 
define the wave number $~k~$ for the interacting system by putting
\begin{equation}
E_n~=~{\hbar^2 k^2 \over 2\mu_{ab}}~.
\label{en}
\end{equation}
Let $~g(k)\,dk~$ be the number of states with wave number lying 
between $~k~$ and $~k+dk~$, and let  $~g^{(0)}(k)\,dk~$ denote 
the corresponding quantity for the noninteracting system. 
Then Eq.~(\ref{defZab}) can be written in the form
\begin{equation}
Z_{ab}~=~\sum_B e^{-\beta E_B}
+\int_0^\infty dk\,\{g(k)-g^{(0)}(k)\}\,
\exp\left(-\beta{\hbar^2 k^2 \over 2\mu_{ab}}\right)
\label{Zab_1}
\end{equation}
The difference in density of states is related to the scattering 
phase shifts by the following argument.\cite{SS73D75}
The relative wave function can be factorized 
[see Eq.~(\ref{vpm2})] into a product of a  trivial azimuthal 
part and non-trivial radial wave function $~R_m(\rho)$, 
which satisfies Eq.~(\ref{vpm3}).
For large value of $~\rho~$ where the potential is assumed 
negligible, 
\begin{equation}
R_m(\rho\rightarrow\infty)~\propto~
\cos\{k\rho-(2m+1)\pi/4+\delta_m(k)\}~,
\label{Rmk_int}
\end{equation}
which defines the phase shift $~\delta_m(k)~$ of the $m$-th 
partial wave. 
For the noninteracting system all the phase shifts 
$~\delta_m(k)\equiv 0$.
If the system is placed within a circle of radius $~R$, the 
vanishing of the wave function at the boundary requires 
that the allowed values of $~k~$ are given by  
\begin{equation}
kR\,-\,(2m+1)\pi/4\,+\,\delta_m(k)~=~\left(n+{1\over 2}\right)\pi~,
\label{kRint}
\end{equation}
for the interacting system, and
\begin{equation}
kR\,-\,(2m+1)\pi/4~=~\left(n+{1\over 2}\right)\pi~,
\label{kRnonint}
\end{equation}
for the noninteracting system, where $~n\,=\,0,\,1,\,2,\ldots\,$.
For a given $~m$, changing $~n~$ by one unit causes $~k~$ to change 
by the respective amounts $~\Delta k~$ and $~\Delta k^{(0)}$:
\begin{eqnarray}
\Delta k~&=&~{\pi \over R\,+\,(d\delta_m(k)/dk)}~,\\
\Delta k^{(0)}~&=&~{\pi \over R}~.
\label{deltak}
\end{eqnarray}
These are the spacings of eigenvalues for a given $~m$. 
Let the number of states of a given $~m~$ with wave number lying 
between $~k~$ and $~k+dk~$ be denoted by $~g_m(k)\,dk~$ and
$~g^{(0)}_m(k)\,dk~$ for the two cases. 
We must have
\begin{eqnarray}
g_m(k)\,\Delta k~&=&~1~,\\
g^{(0)}_m(k)\,\Delta k~&=&~1
\label{gmk_1}
\end{eqnarray}
or
\begin{eqnarray}
g_m(k)~&=&~{1\over\pi}\,\left(R\,+\,{d\delta_m(k) \over dk}\right)~,\\
g^{(0)}_m(k)~&=&~{1\over\pi}\,R~.
\label{gmk_2}
\end{eqnarray}
Therefore
\begin{equation}
g_m(k)\,-\,g^{(0)}_m(k)~=~{1\over\pi}\,{d\delta_m(k) \over dk}~.
\label{gmk_3}
\end{equation}
Summing Eq.~(\ref{gmk_3}) over all allowed $~m~$ we obtain 
\begin{equation}
g(k)\,-\,g^{(0)}(k)~=~{1\over\pi}\,
\sum_m{d\delta_m(k) \over dk}~.
\label{gkdiff}
\end{equation}
Substituting Eq.~(\ref{gkdiff}) into Eq.~(\ref{Zab_1}) yields
\begin{eqnarray}
Z_{eh}&~=~&\sum_{B}e^{-\beta E_B} 
\nonumber\\
&~+~&{1\over\pi}\,\int_0^\infty
{\left(\sum_{m=-\infty}^{\infty}{d\delta_m(k)\over dk}\right)\, 
\exp\left(-\beta {\hbar^2 k^2 \over 2\mu_{ab}}\right)\,dk}~,
\label{bu2D_1}
\end{eqnarray}
which coincides with Eq.~(\ref{eq03}) if we change the notation for 
the bound-state energy from $~E_B~$ to $~E_{m,\nu}$, where subscript 
$~\nu~$ enumerates bound states with a given $~m$. 

Now, having evaluated the interaction part of the partition function,
we can obtain the coefficient $~{\cal C}_{1,1}~$ needed in the density 
expansion, by substituting Eq.~(\ref{Q11viaZab}) into Eq.~(\ref{c11_1}) 
and taking into account Eq.~(\ref{Q11eval}), 
\begin{equation}
{\cal C}_{1,1}~=~g_a\,g_b\,{1\over\lambda_{M_a+M_b}^2}Z_{ab}~=~
\lambda^2_{\mu_{ab}}\,{g_a\over\lambda^2_{M_a}}\,
{g_b\over\lambda^2_{M_b}}\,Z_{ab}~.
\label{c11_2}
\end{equation}

Up to this point we have considered a system of two distinguishable 
particles (e.g. an electron and a hole).
To calculate the remaining coefficient, $~{\cal C}_{2,0}$, in 
Eq.~(\ref{naapprox}) we must now consider the interaction of
indistinguishable particles. 
For such particles the sum over different $~m~$ is modified, 
e.g., for two spinless bosons the relative wave function must be 
symmetric, this means that only even values of $~m~$ are possible.
For the case of two fermions with the same spin the coordinate wave 
function must by antisymmetric. 
Let us consider a case of the two fermions or bosons of the same type 
$~a~$, with the quantum state degeneracy $~g_a$. 
This degeneracy is usually associated with the particle spin $~s$, 
so that $~g_a=2s+1$. 
For a two-particle system there are $~g_a^2=(2s+1)^2~$ spin states of 
which a fraction, $~(s+1)/(2s+1)=(g_a+1)/(2g_a)$, are symmetric and 
$~s/(2s+1)=(g_a-1)/(2g_a)~$ are antisymmetric. 
For a fermion system, the symmetric spin states must be multiplied 
by antisymmetric spatial states (odd $~m$) and the antisymmetric spin 
states multiply symmetric spatial states (even $~m$). 
A similar argument is applicable for bosons.
Thus, for the 2D system of interacting particles of the type $~a$, 
the two-body partition function is 
\begin{equation}
{\cal Q}_{2,0}(A,T)~=~{\cal Q}^{(0)}_{2,0}(A,T)\,+\,
g_a^2\,{A\over\lambda^2_{M_a}}\,Z_{aa}~,
\label{Q20viaZaa}
\end{equation}
where $~Z_{aa}~$ is defined by
\begin{eqnarray}
Z_{aa}~&=&~\sum_{m,\nu}{g_a\,\pm\,(-1)^m \over g_a}\,e^{-\beta E_{m,\nu}}
\nonumber\\ 
&+&{1\over\pi}\,\sum_{m=-\infty}^\infty{g_a\,\pm\,(-1)^m\over g_a}
\int_0^\infty {d\delta_m(k) \over dk}\, 
\exp\left(-\beta {\hbar^2 k^2 \over M_a}\right)\,dk~,
\label{Zaagen}
\end{eqnarray}
Here and in what follows the upper sign is for bosons and the lower 
sign is for fermions.
For the case of repulsive fermions (which do not form bound states) 
and for $~g_a=2~$, Eq.~(\ref{Zaagen}) reduces to Eq.~(\ref{eq04}).

Let us now calculate the two-body partition function of noninteracting 
bosons or fermions, $~{\cal Q}^{(0)}_{2,0}$, which is given by
\begin{eqnarray}
{\cal Q}^{(0)}_{2,0}(A,T)~&=&~{A\over\lambda^2_{2M_a}}\,{g_a\over2}\,
\int d^2\bbox{\rho}{\sum_{{\bf k}}}'\Bigl\{
(g_a\pm 1)\,|\psi_s^{(0)}({\bf k},{\bbox{\rho}})|^2
\nonumber\\
&+&~(g_a\mp 1)\,|\psi_a^{(0)}({\bf k},{\bbox{\rho}})|^2\Bigr\}\,
\exp\left(-\beta {\hbar^2 k^2 \over M_a}\right)\,dk~,
\label{Q20ni_1}
\end{eqnarray}
where $~\psi_s^{(0)}({\bf k},\bbox{\rho})~$ and 
$~\psi_a^{(0)}({\bf k},\bbox{\rho})~$ are symmetric and antisymmetric 
eigenfunctions of the noninteracting Hamiltonian of the relative motion, 
i.e.,
\begin{equation}
\psi_s^{(0)}({\bf k},\bbox{\rho})~=~\sqrt{{2\over A}}\,
\cos({\bf k}\cdot\bbox{\rho})
\label{psi0s}
\end{equation}
and
\begin{equation}
\psi_a^{(0)}({\bf k},\bbox{\rho})~=~\sqrt{2 \over A}\,
\sin({\bf k}\cdot\bbox{\rho})~.
\label{psi0a}
\end{equation}
The summation in Eq.~(\ref{Q20ni_1}) must be performed over all
{\it different} two-particle states.
Wave vectors $~{\bf k}~$ and $~-{\bf k}~$ correspond to the same state
(${\bf k}~$ and $~-{\bf k}~$ transfer into each other upon exchange 
of the two indistinguishable particles), this state should not be 
counted twice and
\begin{equation}
{\sum_{{\bf k}}}'
~\longrightarrow~{1\over2}\,{A\over (2\pi)^2}\,
\int d^2 {\bf k}~.
\label{sumk}
\end{equation}
Then $~{\cal Q}^{(0)}_{2,0}~$ can be written as
\begin{equation}
{\cal Q}^{(0)}_{2,0}(A,T)~=~g_a\,{A\over\lambda^2_{M_a}}\,
\Bigl\{(g_a \pm 1){\cal I}_s\,+\,(g_a \mp 1){\cal I}_a\Bigr\}~,
\label{Q20ni_2}
\end{equation}
where
\begin{eqnarray}
{\cal I}_s~&=&~{1\over (2\pi)^2}\,\int{d^2\bbox{\rho}\,d^2{\bf k}\,
\cos^2({\bf k}\cdot\bbox{\rho})
\exp\left(-\beta {\hbar^2 k^2 \over M_a}\right)}
\nonumber\\
&=&~{1\over2}\,\left({A\over2\lambda^2_{M_a}}\,
+\,{1\over 4}\right)
\label{Is}
\end{eqnarray}
and 
\begin{eqnarray}
{\cal I}_a~&=&~{1\over (2\pi)^2}\,\int{d^2\bbox{\rho}\,d^2{\bf k}\,
\sin^2({\bf k}\cdot\bbox{\rho})
\exp\left(-\beta {\hbar^2 k^2 \over M_a}\right)}
\nonumber\\
&=&~{1\over2}\,\left({A\over2\lambda^2_{M_a}}\,
-\,{1\over 4}\right)~.
\label{Ia}
\end{eqnarray}
Finally,
\begin{equation}
{\cal Q}^{(0)}_{2,0}(A,T)~=~{1\over 2}\,
\left(g_a{A\over\lambda^2_{M_a}}\right)^2
~\pm~{1\over 4}\,g_a{A\over\lambda^2_{M_a}}
~=~{1\over 2}{\cal Q}^2_{1,0}
~\pm~{1\over 4}\,g_a{A\over\lambda^2_{M_a}}~.
\label{Q20ni_fin}
\end{equation}

Substituting Eqs.~(\ref{Q20viaZaa}) and (\ref{Q20ni_fin}) into 
Eq.~(\ref{c20_1}) yields
\begin{equation}
{\cal C}_{2,0}~=~{1\over 2}\,\lambda^2_{\mu_{aa}}
\left({g_a\over\lambda^2_{M_a}}\right)^2\,Z_{aa}
~\pm~{1\over 4}\,{g_a\over\lambda^2_{M_a}}~.
\label{c20_2}
\end{equation}

Substituting Eqs.~(\ref{c10_2}), (\ref{c11_2}), and (\ref{c20_2})
into Eq.~(\ref{naapprox}) we find the following expression for 
the total density of the component $~a$:
\begin{eqnarray}
n_a~&\approx&~{g_a \over \lambda^2_{M_a}}\,z_a
~\pm~{1\over 2}\,{g_a\over\lambda^2_{M_a}}\,z_a^2
\nonumber\\
&+&\lambda_{\mu_{aa}}^2 
\left({g_a\over\lambda^2_{M_a}}\right)^2\,Z_{aa}\,z_a^2
~+~\lambda^2_{\mu_{ab}}\,{g_a\over\lambda^2_{M_a}}\,
{g_b\over\lambda^2_{M_b}}\,Z_{ab}\,z_a\,z_b~.
\label{ntota_a}
\end{eqnarray}
The first two terms in Eq.~(\ref{ntota_a}) do not depend on 
interparticle interaction (however the second term depends 
via its sign on the statistics of the particles). 
It is natural to call the sum of two first terms in 
Eq.~(\ref{ntota_a}) as {\it uncorrelated} density, 
we denote it as $~n_a^0$.
Note that these terms are the first two terms in the low-density 
expansion of the well-known expression\cite{HKbook} for the density 
of the noninteracting 2D Bose or Fermi gases:
\[
n_a^0~=~\mp{g_a \over \lambda_{M_a}^2}\,\ln(1\mp z_a)~.
\]
The sum of the last two terms in Eq.~(\ref{ntota_a}) is the 
interaction-dependent {\it correlated} density $~n_a^{corr}$.

In the low-density limit the second term in Eq.~(\ref{ntota_a}) 
is much smaller than the first one, and within the same accuracy 
as Eq.~(\ref{ntota_a}) we can write 
\begin{equation}  
n_a^{corr}~\approx~\sum_{b}n^0_a n^0_b\,\lambda^2_{\mu_{ab}}\,Z_{ab}~.
\label{lawmass}
\end{equation}
Equation~(\ref{lawmass}) coincides with Eq.~(\ref{eq02}),
and it constitutes the modified law of the mass action in two 
dimensions.

\newpage

\newpage

\newpage

\newpage

\newpage

\newpage

\begin{references}
\bibitem{m1note1} For $~m=1$, when a half-bound state occurs, $~\nu_m~$ 
denotes the number of bound states plus one. The wave function of the 
half-bound state tends to zero at infinity, although it does not decay 
fast enough at infinity to be square integrable. 
More on this subject can be find in Refs.~\onlinecite{Bolle,Dong}.
\bibitem{Surprise} Surprisingly, the situation changes drastically 
in the zero-density limit ($q_sa^*\to 0$, while $~k_BT/{\rm Ry}^*\sim 1$). 
It can be shown that in this limit $~Z_{eh}+Z_{ee}=
N_b+f(k_BT/{\rm Ry}^*,\,q_sa^*)$, 
where $~|f(k_BT/{\rm Ry}^*,\,q_sa^*)| \ll N_b$, $~N_b~$ being the number 
of bound states. 
Then, using Eqs.~(\ref{modsemalpha}) and (\ref{nb3}), we get $~\alpha\to2/3$, 
when $~q_sa^*\to 0$, whereas a simple, single-bound-state law of 
the mass action gives $~\alpha \to 1~$ in the same limit.
\end{references}
\end{document}